\documentclass[a4paper,12pt]{article}
\usepackage{epsfig}
\usepackage{citesort}
\date{\today}
 \normalsize

\newcommand{\bmat}{\left(\begin{array}}
\newcommand{\emat}{\end{array}\right)}
\newcommand{\be}{\begin{equation}}
\newcommand{\ee}{\end{equation}}
\newcommand{\bea}{\begin{eqnarray}}
\newcommand{\eea}{\end{eqnarray}}

\def\lsim{\raise0.3ex\hbox{$\;<$\kern-0.75em\raise-1.1ex\hbox{$\sim\;$}}}
\def\gsim{\raise0.3ex\hbox{$\;>$\kern-0.75em\raise-1.1ex\hbox{$\sim\;$}}}
\def\Frac#1#2{\frac{\displaystyle{#1}}{\displaystyle{#2}}}

\def\arnps#1#2#3{{\it Annu. Rev. Nucl. Part. Sci.\/} {\bf#1} (#2) #3}

\def\ibid#1#2#3{\emph{ibid.} {\bf #1} (#2) #3}

\def\npb#1#2#3{{\it Nucl.~Phys.\/}~{\bf B #1} (#2) #3}
\def\npps#1#2#3{{\it Nucl.~Phys.~(Proc.~Suppl.)\/}~{\bf #1} (#2) #3}
\def\plb#1#2#3{{\it Phys.~Lett.\/}~{\bf B #1} (#2) #3}

\def\prd#1#2#3{{\it Phys.~Rev.\/}~{\bf D#1} (#2) #3}
\def\prl#1#2#3{{\it Phys.~Rev.~Lett.\/}~{\bf #1} (#2) #3}

\def\hpph#1{{\tt hep-ph/#1}}

\def\jhep#1#2#3{{\it J.~High Energy Phys.}~{\bf #1} (#2) #3}
\begin{document}
\renewcommand{\thefootnote}{\fnsymbol{footnote}}
\rightline{IPPP/02/36} \rightline{DCPT/02/72}
\rightline{HIP-2002-29/TH} 
\vspace{.3cm} 
{\Large
\begin{center}
{\bf Constraining supersymmetric models from $B_d - \bar{B}_d$ mixing and the 
$B_d \to J/\psi K_S$ asymmetry}
\end{center}}
\vspace{.3cm}

\begin{center}
E. Gabrielli$^{1}$ and S. Khalil$^{2,3}$\\
\vspace{.3cm}
$^1$\emph{
Helsinki Institute of Physics,
     POB 64,00014 University of Helsinki, Finland}
\\
$^2$ \emph{IPPP, Physics Department, Durham University, DH1 3LE,
Durham,~~U.~K.}
\\
$^3$ \emph{Ain Shams University, Faculty of Science, Cairo, 11566,
Egypt.}

\end{center}

\vspace{.3cm}
\hrule \vskip 0.3cm
\begin{center}
\small{\bf Abstract}\\[3mm]
\end{center}
We analyze the chargino contributions to $B_d - \bar{B}_d$ mixing 
and CP asymmetry of the $B_d \to J/\psi K_S$ decay, in the framework of
the mass insertion approximation. We derive model independent bounds on 
the relevant mass insertions. Moreover, we study these contributions in 
supersymmetric models with minimal flavor violation, Hermitian flavor 
structure, and small CP violating phases and universal strength Yukawa
couplings. We show that in supersymmetric models with large flavor mixing, 
the observed values of $\sin 2 \beta$ may be entirely due to the 
chargino--up--squark loops. 

\begin{minipage}[h]{14.0cm}
\end{minipage}
\vskip 0.3cm \hrule \vskip 0.5cm
%%%%%%%%%%%%%%%%%%%%%%%%%%%%%%%%%%
%
\section{{\large \bf Introduction}}
Since its discovery in 1964 in the $K$-meson decays, the origin of CP violation
remains an open question in particle physics. 
In standard model (SM), the phase of 
the Cabbibo--Kobayashi--Maskawa (CKM) quark mixing matrix provides 
an explanation
of the CP violating effect in these decays.
Although the SM is able to account for the observed CP violation in 
the kaon system
and the recent measurement of the (time-dependent) 
CP asymmetry in $B_d \to J/\psi K_S$ 
decays, new
CP violating sources are necessarily required to describe the observed baryon 
asymmetry \cite{bau}. Moreover, it is expected that with $B$ factories, 
the $B$--system will represent an ideal framework for crucial tests of the CP 
violation in SM and probing new physics effects at low energy. 

It is a common feature for any new physics beyond the SM to possess additional
CP violating phases beside the $\delta_{CKM}$ phase. In 
supersymmetric (SUSY) models,
the soft SUSY breaking terms contain several parameters that may be complex, as
may also the SUSY preserving $\mu$ parameter. These new phases have significant
implications for the electric dipole moment (EDM) of electron, neutron, and 
mercury atom~\cite{bound}.
It was shown that the EDM can be suppressed in 
SUSY models with small CP phases ~\cite{phase:edm,savoy}
or in SUSY models with flavor off--diagonal CP 
violation~\cite{phase:edm,hermitian}.

The idea of having small CP phases $(\lsim 10^{-2})$ as an approximate
CP--symmetry at low energy, could be an 
interesting possibility if supported by a mechanism of CP--symmetry 
restoration at high energy scale. However, this mechanism
might also imply that the $\delta_{CKM}$ phase is small~\cite{nir}. 
The large asymmetry of the $B$-meson decay $a_{J/\psi K_S}$ observed 
by BaBar and Belle \cite{babar} 
are in agreement with SM predictions for large $\delta_{CKM}$, 
and thus the idea of small 
phases might be disfavored.

However, in Ref.\cite{branco} it was shown that
in the framework of the minimal supersymmetric standard model (MSSM) with 
non--universal soft terms and large flavor mixing in the Yukawa, 
supersymmetry can give the leading contribution to $a_{J/\psi K_S}$
with simultaneous account for the experimental results in $K$--system. Thus
the supersymmetric models with small CP violating phases at high energy scale
are still phenomenological viable.
An alternative possibility for suppressing the EDMs is that SUSY CP phases
has a flavor off--diagonal character as in the SM~\cite{hermitian,susycp}.
Such models would allow for phases  of order 
${\cal O}(1)$ which may have significant effect in $B$-physics 
\cite{hermitian}.  

A useful tool for analyzing SUSY contributions to 
flavour changing neutral current processes (FCNC)
is provided, as known, by the mass insertion method
\cite{gluino}.
One chooses a basis for the fermion and sfermion states where all the 
couplings of these particles to neutral gauginos are flavour 
diagonal, leaving all the sources of FC inside the off-diagonal terms 
of sfermion mass matrix. These terms are denoted by $(\Delta^{q}_{AB})^{ij}$,
where as usual $A,B=(L,R)$ and $i,j=1,3$ indicate chiral and flavour indices 
respectively and $q=u,d$. The sfermion propagator
is then expanded as a series of $(\delta^{q}_{A B})_{ij}=
(\Delta^{q}_{AB})^{ij}/\tilde{m}^2$, where $\tilde{m}^2$
is an average sfermion mass.
This method allows to parametrize, in a 
model independent way, the main sources of flavour violations in SUSY models.
In this framework, the gluino and 
chargino contributions to the $K$--system have been
analyzed in references \cite{gluino} and \cite{khalil} respectively.
These analyses showed that the bounds 
on imaginary parts of mass insertions, coming from gluino exchanges 
to $\varepsilon_K$ and $\varepsilon'/\varepsilon$, 
are very severe \cite{gluino}, while the corresponding 
ones from chargino exchanges are less constrained \cite{khalil}.
In particular, in order to saturate $\varepsilon_K$ from the 
gluino contributions one should have \cite{gluino} 
$\sqrt{\vert \mathrm{Im}(\delta_{12}^d)^2_{LL} \vert}\sim 10^{-3}$ or
$\sqrt{\vert \mathrm{Im}(\delta_{12}^d
)^2_{LR}\vert}\sim 10^{-4}$, and $\sqrt{\vert \mathrm{Im}(\delta_{12}^d
)^2_{LR}\vert}\sim 10^{-5}$ from $\varepsilon'/\varepsilon$, while 
chargino contributions require \cite{khalil}
$\mathrm{Im}(\delta^u_{12})^2_{LL} \sim 10^{-2}$, for
average squark masses of order of $500$ GeV and gluino masses of the
same order.

Recently, in the framework of mass insertion approximation, 
gluino contributions to the $B_d-\bar{B}_d$ mixing 
and CP asymmetry in the decay $B_d \to J/\psi K_S$
have been analyzed by including next-to-leading order
(NLO) QCD corrections \cite{gluinoB} (see also Ref.\cite{chua}).
However, an analogous study for chargino contributions to 
these processes is still missing. 
This kind of analysis would be interesting for the following reasons.
First, it would provide a new set of upper bounds 
on the mass insertion parameters, namely 
$(\delta^u_{ij})_{AB}$ in the up--squark sector, which  are complementary 
to the ones obtained from gluino exchanges (which only constrain
$(\delta^d_{ij})_{AB}$).
Second, upper bounds on $(\delta^u_{ij})_{AB}$ would be very useful in 
order to perform easy tests on SUSY models which receive from 
chargino exchanges the main contributions to $B_d-\bar{B}_d$ and CP asymmetry.
Indeed, in many SUSY scenarios the gluino exchanges are always sub-leading.

In this paper we focus on the dominant 
chargino contributions to the $B_d-\bar{B}_d$ mixing 
and CP asymmetry $a_{J/\psi K_S}$. We use 
the mass insertion method and derive the corresponding bounds on the
relevant mass insertion parameters. We perform this analysis at the 
NLO accuracy in QCD by using the results available in Ref.\cite{gluinoB}.
As an application of our analysis, 
we also provide a comparative study for 
supersymmetric models with minimal flavor violation, Hermitian flavor 
structure with small CP violating phases and universal strength of 
Yukawa couplings.
We show that in all these scenarios,
by comparing $(\delta^u_{ij})_{AB}$ and $(\delta^d_{ij})_{AB}$ 
with their corresponding upper bounds, the chargino contributions are 
dominant over the gluino ones. 

The paper is organized as follows. In section 2 we present the supersymmetric
contributions to $B_d - \bar{B}_d$ mixing and CP asymmetry
$a_{J/\psi K_S}$. We start with a brief review 
on gluino contributions and then we present our results for the chargino 
ones, both in the mass insertion approach. 
In section 3 we derive model 
independent bounds on the relevant mass insertions involved 
in the $B_d -\bar{B}_d$ mixing 
and $a_{J/\psi K_S}$.
In section 4
we generalize these results by including the case of a light stop--right.
Section 5 is devoted to the study of the 
supersymmetric contribution to $a_{J/\psi K_S}$ in three different
supersymmetric models. We show that the observed values of $\sin 2 \beta$ may 
be entirely due to the chargino---up--squark loops in some classes 
of these models.
Our conclusions are presented in section 6. 
 
%%%%%%%%%%%%%%%%%%%%%%%%%%%%%%%%%%%%%%%%%%%%%%%%%%%%%%%%%%%%%%
\section{{\large \bf Supersymmetric contributions to $\Delta B=2$
transitions}}
We start this section by summarizing the main results on 
$B_d - \bar{B}_d$ mixing and CP asymmetry
$a_{J/\psi K_S}$, then we will consider the
relevant SUSY contributions to the effective Hamiltonian 
for $\Delta B=2$ transitions, given by the chargino and gluino box diagram
exchanges.

In the $B_d$ and  $\bar{B}_d$ system, the flavour eigenstates are given by 
$B_d = (\bar{b} d)$ and $\bar{B}_d = (b \bar{d})$. It is 
customary to denote the corresponding 
mass eigenstates by $B_H = p B_d + q \bar{B}_d$
and $B_L = p B_d - q \bar{B}_d$ where indices H and L 
refer to heavy and light mass eigenstates
respectively, and 
$ p=(1+\bar{\varepsilon}_B)/\sqrt{2(1+|\bar{\varepsilon}_B|)}, ~~
q=(1-\bar{\varepsilon}_B)/\sqrt{2(1+|\bar{\varepsilon}_B|)}$
where $\bar{\varepsilon}_B$ is the corresponding 
CP violating parameter in the $B_d -\bar{B}_d$ system, analogous of
$\bar{\varepsilon}$ in the kaon system \cite{buras1}.
Then the strength of $B_d - \bar{B}_d$ mixing is described by
the mass difference 
\be
\Delta M_{B_d} = M_{B_H} - M_{B_L}.
\ee
whose present experimental value is 
$\Delta M_{B_d} = 0.484 \pm 0.010~\mathrm{(ps)}^{-1}$~\cite{buras1}.

The CP asymmetry of the  $B_d$ and $\bar{B}_d$ meson decay to the CP 
eigenstate $\psi K_S$ is given by
\be 
a_{\psi K_S}(t) = \frac{\Gamma(B^0_d(t) \to \psi K_S) -
\Gamma(\bar{B}^0_d(t) \to \psi K_S)}{\Gamma(B^0_d(t) \to \psi K_S) +
\Gamma(\bar{B}^0_d(t) \to \psi K_S)}=- a_{\psi K_S} \sin(\Delta
m_{B_d} t).
\ee
The most recent measurements of this asymmetry are given by~\cite{babar}
\bea
a_{\psi K_S}&=& 0.59 \pm 0.14 \pm 0.05 \quad \quad
\mathrm{(BaBar)} \;, \nonumber \\
a_{\psi K_S}&=& 0.99 \pm 0.14 \pm 0.06 \quad \quad
\mathrm{(Belle)} \;.
\label{2beta:bounds}
\eea
where the second and third numbers correspond to statistic and systematic 
errors respectively, and so the present world average is given by 
$a_{\psi K_S}=0.79\pm 12$.
These results show that there is a large CP asymmetry 
in the $B$ meson system. This implies that either the CP is not an approximate 
symmetry in nature and that the CKM mechanism is the dominant source of 
CP violation~\cite{hermitian} or CP is an approximate symmetry with large 
flavour structure beyond the standard CKM matrix~\cite{branco}. 
Generally, $\Delta M_{B_d}$ and $a_{\psi K_S}$
can be calculated via
\bea
&&\Delta M_{B_d} = 2 
\vert \langle B^0_d \vert H_{\mathrm{eff}}^{\Delta B=2} \vert 
\bar{B}^0_d \rangle \vert \;, 
\label{DeltaMB}
\\
&&a_{\psi K_S} = \sin 2 \beta_{\mathrm{eff}} \;,~ \mathrm{and}~~~~ 
\beta_{\mathrm{eff}} = \frac{1}{2}\mathrm{arg} \langle B^0_d \vert 
H_{\mathrm{eff}}^{\Delta B=2} \vert \bar{B}^0_d \rangle \;.
\label{sin2b}
\eea
where $H_{\mathrm{eff}}^{\Delta B=2}$ is the effective Hamiltonian 
responsible of the ${\Delta B=2}$ transitions.
In the framework of the standard model (SM), 
$a_{\psi K_S}$ can be easily related 
to one of the inner angles of the unitarity triangles and 
parametrized by the $V_{CKM}$ elements as follows
\be
a_{\psi K_S}^{\mathrm{SM}} = \sin 2 \beta,~~ \beta = \mathrm{arg}\left(-\frac{V_{cd}
V_{cb}^*}{V_{td} V_{tb}^*} \right),
\ee

In supersymmetric theories the effective Hamiltonian 
for $\Delta B= 2$ transitions, can be generated,
in addition to the $W$ box diagrams of SM,
through other box diagrams mediated by charged Higgs, neutralino, photino, 
gluino, and chargino exchanges. The 
Higgs contributions are suppressed by the quark masses and can be neglected. 
The neutralino and photino exchange diagrams are also very suppressed compared 
to the gluino and chargino ones, due to their electroweak
neutral couplings to fermion and sfermions.
Thus, the dominant SUSY contributions to the 
off diagonal entry in the $B$-meson mass matrix, $\mathcal{M}_{12}(B_d) = 
\langle B^0_d \vert H_{\mathrm{eff}}^{\Delta B=2} \vert \bar{B}^0_d \rangle$,
is given by
\begin{equation}
\mathcal{M}_{12}(B_d) = \mathcal{M}_{12}^{\mathrm{SM}}(B_d) + 
\mathcal{M}_{12}^{\tilde{g}}(B_d) + \mathcal{M}_{12}^{\tilde{\chi}^+}(B_d).
\end{equation}
where $\mathcal{M}_{12}^{\mathrm{SM}}(B_d)$, 
$\mathcal{M}_{12}^{\tilde{g}}(B_d)$, and $\mathcal{M}_{12}^{\tilde{\chi}^+}$
indicate the SM, gluino, and chargino contributions respectively.
The SM contribution 
is known at NLO accuracy in QCD \cite{buras1} 
(as well as the leading SUSY contributions \cite{gluinoB})
and it is given by
\be
\mathcal{M}_{12}^{\mathrm{SM}}(B_d)= \frac{G_F^2}{12 \pi^2} \eta_{B} 
\hat{B}_{B_d} f^2_{B_d}
M_{B_d} M_W^2 (V_{td} V_{tb}^*)^2 S_0(x_t), 
\ee
where $f_{B_d}$ is the B meson decay constant, $\hat{B}_{B_d}$ is
the renormalization group invariant $B$ parameter (for its
definition and numerical value, see Ref. \cite{buras1}
and reference therein) and $\eta=0.55\pm 0.01$. The function $S_0(x_t)$,
connected to the $\Delta B=2$ box diagram with $W$ exchange, is given by
\be
S_0(x_t) = \frac{4 x_t - 11 x_t^2 +x_t^3}{4(1-x_t)^2} - 
\frac{3 x_t^3 \ln x_t}{2(1-x_t)^3},
\ee
where $x_t= M_t^2/M_W^2$. 

The effect of supersymmetry can be simply
described by a dimensionless parameter $r_d^2$ and a phase $2 \theta_d$
defined as follows
\begin{equation}
r_d^2 e^{2 i \theta_d} = \frac{\mathcal{M}_{12}(B_d)}{\mathcal{M}_{12}^{\mathrm{SM}}
(B_d)},
\end{equation}
where 
$\Delta M_{B_d} = 2 \vert \mathcal{M}_{12}^{\mathrm{SM}}(B_d)\vert r_d^2$.
Thus, in the presence of SUSY contributions, the
CP asymmetry $B_d \to \psi K_S$ is modified, and now we have 
\be
a_{\psi K_S} = \sin 2 \beta_{\mathrm{eff}} = \sin (2 \beta + 2 \theta_d)\;.
\ee
Therefore, the measurement of $a_{\psi K_S}$ would not determine $\sin 2 \beta$ but 
rather $\sin 2 \beta_{\mathrm{eff}}$, where
\begin{equation}
2 \theta_d = \mathrm{arg}\left(1+\frac{\mathcal{M}_{12}^{\mathrm{SUSY}}(B_d)}
{\mathcal{M}_{12}^{\mathrm{SM}}(B_d)}\right),
\end{equation}
and $\mathcal{M}_{12}^{\mathrm{SUSY}}(B_d)= \mathcal{M}_{12}^{\tilde{g}}(B_d) + 
\mathcal{M}_{12}^{\tilde{\chi}^+}(B_d)$. 
%%%%%%%%%%%%%%%%%%%%%%%%%%%%%%%%%%%%%%%%%%%%%%%%%%%%%%%%%%%%%%

\subsection{{\large \bf Gluino contributions}}
The most general effective Hamiltonian for $\Delta B=2$ 
processes, induced by gluino and chargino exchanges through 
$\Delta B=2$ box diagrams, can be expressed as
\be
H^{\Delta B=2}_{\rm eff}=\sum_{i=1}^5 C_i(\mu) Q_i(\mu) + 
\sum_{i=1}^3 \tilde{C}_i(\mu) \tilde{Q}_i(\mu) + h.c. \;,
\label{Heff}
\ee
where $C_i(\mu)$, $\tilde{C}_i(\mu)$ and $Q_i(\mu)$, $\tilde{Q}_i(\mu)$
are the Wilson coefficients and operators respectively 
renormalized at the scale $\mu$, with
\bea 
Q_1 &=&\bar{d}_L^{\alpha} \gamma_{\mu} b_L^{\alpha}~ 
\bar{d}_L^{\beta} \gamma_{\mu} b_L^{\beta},~~~~
Q_2 = \bar{d}_R^{\alpha} b_L^{\alpha}~ 
\bar{d}_R^{\beta} b_L^{\beta},~~~~ Q_3 = \bar{d}_R^{\alpha} b_L^{\beta}~ 
\bar{d}_R^{\beta} b_L^{\alpha},\nonumber\\
Q_4 &=& \bar{d}_R^{\alpha} b_L^{\alpha}~ \bar{d}_L^{\beta} 
b_R^{\beta},~~~~~~~~~
Q_5 = \bar{d}_R^{\alpha} b_L^{\beta}~ \bar{d}_L^{\beta} b_R^{\alpha}\, .
\label{operators}
\eea
In addition, the operators $\tilde{Q}_{1,2,3}$ are obtained from $Q_{1,2,3}$ 
by exchanging $L \leftrightarrow R$. 

Now we summarize the main results for gluino contributions
to the above Wilson coefficients at SUSY scale, in the 
framework of mass insertion approximation.
As we will show in the next section, in order to connect 
the Wilson coefficients at SUSY scale with the corresponding ones at 
low energy scale $\mu\simeq {\cal O}(m_b)$, 
the RGE equations for QCD corrections must be solved.
In the case of the gluino exchange all 
the above operators give significant contributions 
and the corresponding Wilson coefficients are given by \cite{gluino},
\cite{gluinoB} 
\bea
C_1(M_S) &=& -\frac{\alpha_s^2}{216 m_{\tilde{q}}^2} \left(24x f_6(x)+
66\tilde{f}_6(x)
\right) (\delta_{13}^d)^2_{LL},\nonumber\\
C_2(M_S) &=& -\frac{\alpha_s^2}{216 m_{\tilde{q}}^2} 204 x  f_6(x) 
(\delta_{13}^d)^2_{RL},\nonumber\\
C_3(M_S) &=& \frac{\alpha_s^2}{216 m_{\tilde{q}}^2} 36 x  f_6(x) 
(\delta_{13}^d)^2_{RL},\\
C_4(M_S) &=& -\frac{\alpha_s^2}{216 m_{\tilde{q}}^2} \left[\left(504 x f_6(x) - 72 
\tilde{f}_6(x)\right) (\delta_{13}^d)_{LL}(\delta_{13}^d)_{RR}
-132 \tilde{f}_6(x) (\delta_{13}^d)_{LR}(\delta_{13}^d)_{RL}\right] ,
\label{wilsong}\nonumber\\
C_5(M_S) &=& -\frac{\alpha_s^2}{216 m_{\tilde{q}}^2} \left[\left(24 x f_6(x) +120 
\tilde{f}_6(x)\right) (\delta_{13}^d)_{LL}(\delta_{13}^d)_{RR}
-180 \tilde{f}_6(x) (\delta_{13}^d)_{LR}(\delta_{13}^d)_{RL}\right],\nonumber
\eea
where $x=m^2_{\tilde{g}}/m^2_{\tilde{q}}$ and $\tilde{m}^2$ is 
an average squark mass.
The expression for the 
functions $f_6(x)$ and $\tilde{f}_6(x)$ can be found in Ref.\cite{gluinoB}. 
The Wilson coefficients 
$\tilde{C}_{1-3}$ are simply obtained by 
interchanging $L\leftrightarrow R$ in the mass insertions appearing in 
$C_{1-3}$. 
%%%%%%%%%%%%%%%%%%%%%%%%%%%%%%%%%%%%%%%%%%%%%%%%%%%%%%%%%%%%%%%%%%%%%%%%

\subsection{{\large \bf Chargino contributions}}

Here we present our results for the chargino contributions to the 
effective Hamiltonian in Eq.(\ref{Heff}) in the mass insertion approximation.
The leading diagrams are illustrated in Fig. 1, where the cross in the
middle of the squark propagator represents a single mass insertion. 
As we will explain in more details below,
the dominant chargino exchange, can significantly
affect the operators $Q_1$ and $Q_3$ only.
We remind here that in the case of 
the $K-\bar{K}$ mixing, the relevant chargino exchange affects 
only the operator $Q_1$ \cite{khalil}, as in the SM.
%---------------------------------------
\begin{figure}[t]
\begin{center} 
\hspace*{-7mm}
%\vskip 10cm
\epsfig{file=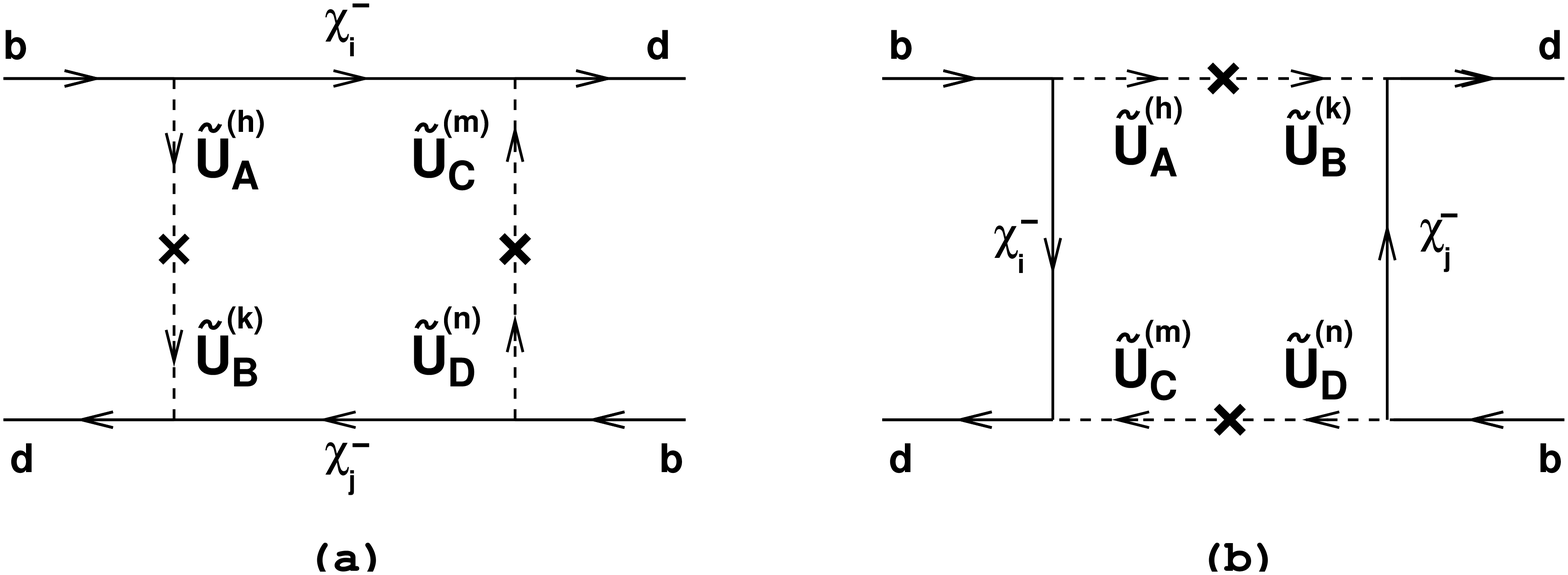,width=13cm,height=5cm}\\
\caption{{\small The leading chargino--up--squark contribution to the 
$B_d - \bar{B}_d$ mixing.}}
\label{fig1}
\end{center}
\end{figure}
%---------------------------------------

In the framework of mass insertion approximation, one chooses
a basis (super-CKM basis) where the couplings of 
the fermions and sfermions to neutral
gauginos are flavour diagonal. 
In this basis, the interacting Lagrangian involving charginos is given by
\bea
\mathcal{L}_{q\tilde{q}\tilde{\chi}^+} &=& - g~ \sum_k \sum_{a,b}~ \Big(~ V_{k1}~ 
K_{ba}^*~ \bar{d}_L^a~ (\tilde{\chi}^+)^*~ \tilde{u}^b_L -  U_{k2}^*~ 
(Y_d^{\mathrm{diag}} . K^+)_{ab}~ \bar{d}_R^a~ (\tilde{\chi}^+)^*~ 
\tilde{u}^b_L \nonumber\\ && - V_{k2}^*~ (K.Y_u^{\mathrm{diag}})_{ab}~
\bar{d}_L^a~ (\tilde{\chi}^+)^*~ \tilde{u}^b_R~ \Big).
\label{vertices}
\eea
where $Y_{u,d}^{\mathrm{diag}}$ are the diagonal Yukawa matrices, and
$K$ is the usual CKM matrix. The indices $a,b$ and $k$ label
flavour and chargino mass eigenstates respectively, and
$V$, $U$ are the chargino mixing matrices defined by
\be
U^* M_{\tilde{\chi}^+} V^{-1} = \mathrm{diag}(m_{\tilde{\chi}_1^+},
m_{\tilde{\chi}_2^+}),~ \mathrm{and}~ M_{\tilde{\chi}^+} = \left(
\begin{array}{cc}
M_2 & \sqrt{2} M_W \sin \beta \\
\sqrt{2} M_W \cos \beta & \mu
\end{array}\right) \;.
\ee
As one can see from Eq.(\ref{vertices}),
the Higgsino couplings are suppressed by Yukawas of the light quarks, and 
therefore they are negligible, except for the stop--bottom
interaction which is directly enhanced by the top Yukawa ($Y_t$). 
The other vertex involving the down and stop 
could also be enhanced by $Y_t$, but one should pay the price of 
a $\lambda^3$ suppression, where $\lambda$ is the Cabibbo mixing.
Since in our analysis we adopt the approximation of
retaining only terms proportional to order $\lambda$, we 
will neglect the effect of this vertex. Moreover, we also set to zero
the Higgsino contributions proportional to Yukawa couplings of light quarks 
with the exception of the bottom Yukawa $Y_b$, 
since its effect could be enhanced by large $\tan{\beta}$.
In this respect, it is clear that the chargino contribution to 
the Wilson coefficients $C_4$ and 
$C_5$ is negligible. Furthermore, due to the colour structure of chargino 
box diagrams there is no contribution to $C_2$ or $\tilde{C}_2$. 
However, as we will show in the next section, 
chargino contributions to $C_2$ or $\tilde{C}_2$ will be always induced 
at low energy by QCD corrections through the mixing with $C_3$.

Now we calculate the relevant Wilson coefficients 
$C_{1,3}^{\chi}(M_S)$ at SUSY scale $M_S$, by 
using the mass insertion approximation. As mentioned in this case the flavor 
mixing is displayed by the non--diagonal entries of the sfermion 
mass matrices. 
Denoting by $\Delta^q_{AB}$ the off--diagonal terms 
in the sfermion $(\tilde{q}=\tilde{u},\tilde{d})$ 
mass matrices for the up, down respectively, 
where $A,B$ indicate chirality couplings to
fermions $A,B=(L,R)$, the A--B squark propagator can be expanded as 
\be
\langle \tilde{q}^a_A \tilde{q}^{b*}_B \rangle = 
i \left(k^2{\bf 1} - \tilde{m}^2 {\bf 1}
- \Delta_{AB}^q\right)^{-1}_{ab} \simeq \frac{i \delta_{ab}}
{k^2 - \tilde{m}^2} + 
~\frac{i (\Delta_{AB}^q)_{ab}}{(k^2 -\tilde{m}^2)^2} + 
{\cal O}(\Delta^2) ,
\ee
where $q=u,d$ selects up, down sector respectively, 
$a,b=(1,2,3)$ are flavour indices, ${\bf 1}$ 
is the unit matrix, and $\tilde{m}$ is the average squark mass.
As we will see in the following, it is convenient to parametrize 
this expansion in terms of the dimensionless quantity 
$(\delta_{AB}^q)_{ab} \equiv (\Delta^q_{AB})_{ab}/\tilde{m}^2$.
At the first order in the mass insertion approximation,  we find 
for the Wilson coefficients $C_{1,3}^{\chi}(M_S)$ the following
result
\bea
C_1^{\chi}(M_S) &=& \frac{g^4}{768 \pi^2 \tilde{m}^2} 
\sum_{i,j}
\Big\{\vert V_{i1}\vert^2 \vert V_{j1} \vert^2\left( 
(\delta^u_{LL})_{31}^2 + 2\lambda (\delta^u_{LL})_{31}
(\delta^u_{LL})_{32}\right)
\nonumber \\
&-& 2 Y_t  
\vert V_{i1}\vert^2 V_{j1} V_{j2}^*\Big(
(\delta^u_{LL})_{31} (\delta^u_{RL})_{31}+\lambda
(\delta^u_{LL})_{32} (\delta^u_{RL})_{31}+\lambda
(\delta^u_{LL})_{31} (\delta^u_{RL})_{32}\Big)
\nonumber \\
&+& Y_t^2 V_{i1} V_{i2}^{*} V_{j1} V_{j2}^{*}\Big(
(\delta^u_{RL})_{31}^2 +2\lambda
(\delta^u_{RL})_{31} (\delta^u_{RL})_{32}~ \Big)\, \Big\} L_2(x_i,x_j),
\label{C1}
\\
C_3^{\chi}(M_S) &=& \frac{g^4\, Y_b^2}{192 \pi^2 \tilde{m}^2}
\sum_{i,j}
U_{i2} U_{j2} V_{j1} V_{i1}~ \Big( (\delta^u_{LL})_{31}^2 + 
2 \lambda  (\delta^u_{LL})_{31} (\delta^u_{LL})_{32} \Big) 
L_0(x_i, x_j),
\label{C3}
\eea
where 
$x_i= m^2_{\tilde{\chi}_i^+}/\tilde{m}^2$, and the functions $L_0(x,y)$
and $L_2(x,y)$ are given by 
\bea
L_0(x,y)&=&\sqrt{xy}\left(\frac{x\, h_0(x)- y\, h_0(y)}{x-y}\right),
~~
h_0(x)=\Frac{-11+7x-2x^2}{(1-x)^3}- \Frac{6 \ln x}{(1-x)^4}
\nonumber \\
L_2(x,y) &=& \Frac{x\, h_2(x) -y\, h_2(y)}{x-y},~~~~~~~~~~~~~
h_2(x)=\Frac{2+5 x -x^2}{(1-x)^3}
+ \Frac{6 x \ln x}{(1-x)^4}
\eea
As in the gluino case, 
the corresponding results for $\tilde{C_1}$ and $\tilde{C_1}$ coefficients 
are simply obtained by interchanging $L\leftrightarrow R$ in 
the mass insertions appearing in the expressions for $C_{1,3}$.
%%%%%%%%%%%%%%%%%%%%%%%%%%%%%%%%%%%%%%%%%%%%%%%%%%%%%%%%%%%%%

\section{{\large \bf Constraints from $\Delta M_{B_d}$ and $\sin 2 \beta$}}

In this section we present our numerical results for 
the bounds on $(\delta^u_{AB})_{ij}$
which come from $\Delta M_{B_d}$ and CP violating
parameter $\sin 2 \beta$.
We start with the chargino contributions
which is found to be the dominant SUSY source in various 
models~\cite{ali,branco,hermitian}. 
We also provide analytical expressions for 
$\Delta M_{B_d}$ and $\sin 2 \beta$ as functions of the mass insertions 
in the Wilson coefficients $C_{i}(M_S)$ of Eqs. (\ref{C1}),
(\ref{C3}). In our calculation we take into account NLO QCD corrections in
both Wilson coefficients and hadronic matrix elements given
in \cite{gluinoB}.

In order to connect $C_i(M_S)$ at SUSY scale $M_S$
with the corresponding low energy ones
$C_i(\mu)$ (where $ \mu\simeq {\cal O}(m_b)$),
one has to solve the renormalization group 
equations (RGE) for the Wilson coefficients 
corresponding to the effective Hamiltonian in (\ref{Heff}).
Then, $C_i(\mu)$ will be related to $C_i(M_S)$
by \cite{gluinoB}
\be
C_r(\mu) = \sum_i \sum_s \left( b_i^{(r,s)} + 
\eta c_{i}^{(r,s)}\right) \eta^{a_i} C_s(M_S),
\label{CWnlo}
\ee
where $M_S>m_t$ and $\eta=\alpha_S(M_S)/\alpha_S(\mu)$.
The values of the coefficients $b_i^{(r,s)}$, $c_i^{(r,s)}$, and 
$a_i$ appearing in (\ref{CWnlo}) can be found in Ref.\cite{gluinoB}.
In our analysis the SUSY scale, 
where SUSY particles are simultaneously integrated out,
is identified with the average squark mass $\tilde{m}$.
By using the NLO results of \cite{gluinoB},
we obtain, for the relevant chargino contributions
\be
C_1(\mu)=x_1(\mu)~ C_1(M_S),~~~C_2(\mu)=x_2(\mu)~ C_3(M_S),~~~
C_3(\mu)=x_3(\mu)~ C_3(M_S),
\label{CWlow}
\ee
while for the other coefficients $C_i(\mu)=0,~~(i=4,5)$.
Numerical values for $x_i(\mu)$, evaluated at $\mu=m_b$,
are shown in table (\ref{table1}) for some representative values of $M_S$.
Notice that the coefficients $b_i^{(23)}$ and 
$c_i^{(23)}$ are different from zero and so
the contribution to $C_2(\mu)$ is radiatively generated 
at NLO by the off-diagonal mixing with $C_3(M_S)$.
For the coefficients $\tilde{C}_{1-3}$
hold the same results as in Eq.(\ref{CWlow}) and in table (\ref{table1}), 
since the corresponding 
$\tilde{b}^{(r,s)}_i$ and $\tilde{c}^{(r,s)}_i$ 
coefficients in Eq.(\ref{CWnlo}) are the same as the ones 
for the evolutions of $C_{1-3}$ \cite{gluinoB}.

The off diagonal matrix elements of the operators $Q_i$ 
are given by \cite{gluinoB}
\bea
\langle B_d \vert Q_1 \vert \bar{B}_d \rangle &=& \frac{1}{3} m_{B_d} f_{B_d}^2
B_1(\mu),\nonumber\\
\langle B_d \vert Q_2 \vert \bar{B}_d \rangle &=& -\frac{5}{24} \left(\frac{m_{B_d}}
{m_b(\mu) + m_d(\mu)} \right)^2 m_{B_d} f_{B_d}^2 B_2(\mu),\nonumber\\
\langle B_d \vert Q_3 \vert \bar{B}_d \rangle &=& \frac{1}{24} \left(\frac{m_{B_d}}
{m_b(\mu) + m_d(\mu)} \right)^2 m_{B_d} f_{B_d}^2 B_3(\mu),
\label{matrixelements}
\\
\langle B_d \vert Q_4 \vert \bar{B}_d \rangle &=& \frac{1}{4} \left(\frac{m_{B_d}}
{m_b(\mu) + m_d(\mu)} \right)^2 m_{B_d} f_{B_d}^2 B_4(\mu),\nonumber\\
\langle B_d \vert Q_5 \vert \bar{B}_d \rangle &=& \frac{1}{12} \left(\frac{m_{B_d}}
{m_b(\mu) + m_d(\mu)} \right)^2 m_{B_d} f_{B_d}^2 B_4(\mu).\nonumber
\eea
The value of $B_1$ has been extensively studied 
on the lattice \cite{lattice1}, but
for the other $B_i$ parameters, they have been 
recently calculated on the lattice by the collaboration 
in Ref. \cite{lattice}. In our
analysis we will use the following central values reported in \cite{gluinoB},
namely $B_1(\mu) = 0.87$, $B_2(\mu)=0.82, B_3(\mu) = 1.02,
B_4(\mu)=1.16$, and $B_5=1.91$. 
%%%%%%%%%%%%%%%%%%%%%%%%%%%%%%%%%%%%%%%%%%%%%%%%%%%
\begin{table}[h]
\begin{center}
\begin{tabular}{|c||c|c|c|}
\hline
$M_S$ & $x_1(\mu)$  & $x_2(\mu)$ & $x_3(\mu)$
\\ \hline
\hline
200 & 0.844 & -0.327  & 0.571 \\
\hline
400 & 0.827 & -0.367  & 0.536 \\
\hline
600 & 0.817 & -0.389  & 0.518 \\
\hline
800 & 0.810 & -0.404  & 0.506  \\
\hline
\end{tabular}
\end{center}
\caption{Numerical values for the coefficients $x_i$ (with $i=1,2,3$) in 
Eq.(\ref{CWlow}) for some representative values of SUSY scale $M_S$, and
evaluated at the low energy scale $\mu=m_b$}
\label{table1}
\end{table}
%%%%%%%%%%%%%%%%%%%%%%%%%%%%%%%%%%%%%%%%%%%%%%%%%%%%%%
The same results of Eq.(\ref{matrixelements})
are also valid for the corresponding operators $\tilde{Q}_i$, with same values 
for the $B_i$ parameters, since strong interactions preserve parity.

Now we start our analysis, by discussing first the dominant chargino
contribution to $\Delta M_{B_d}$. Using 
Eqs.(\ref{DeltaMB}), (\ref{C1})--(\ref{C3}), and 
(\ref{CWlow})--(\ref{matrixelements}), 
we obtain for $\Delta M_{B_d}$ the following result
\bea 
\Delta M_{B_d}&=& \frac{g^4 m_{B_d}f_{B_d}^2}{(48\pi)^2\tilde{m}^2}~
| R + \tilde{R}|
\label{deltaMB}
\\
R&=&
\left(
(\delta^u_{LL})_{31}^2 +2\lambda~(\delta^u_{LL})_{31}(\delta^u_{LL})_{32}
\right) \times \nonumber \\
&&
\Big(
2A_1 x_1(\mu) B_1(\mu) +
A_4 X(\mu)\left(x_3(\mu) B_3(\mu)
-5x_2(\mu) B_2(\mu)\right)
\Big) \nonumber \\
&+&
\left(
(\delta^u_{LL})_{31}\delta^u_{RL})_{31}
 +\lambda \Big(
(\delta^u_{LL})_{31}(\delta^u_{RL})_{32}+
(\delta^u_{LL})_{32}(\delta^u_{RL})_{31}
\Big)
\right)
2 A_2 x_1(\mu) B_1(\mu)\nonumber \\
&+&
\left(
(\delta^u_{RL})_{31}^2
 +2\lambda (\delta^u_{RL})_{31}(\delta^u_{RL})_{32}\right)
2A_3x_1(\mu)B_1(\mu)
\label{R}
\eea
where $\tilde{R}$, which parametrizes the contributions of $\tilde{Q}_{1-3}$
operators, is obtained from $R$ by exchanging $L\leftrightarrow R$
in the mass insertions, 
$X(\mu)=\left(m_{B_d}/(m_b(\mu) + m_d(\mu)) \right)^2$, and
the expressions of $A_i$ are given by
\bea
A_1&=&\sum_{i,j}
\vert V_{i1}\vert^2 \vert V_{j1} \vert^2 L_2(x_i,x_j),~~~~
A_2=Y_t  
\sum_{i,j}\vert V_{i1}\vert^2 V_{j1} V_{j2}^*L_2(x_i,x_j),\nonumber \\
A_3&=&Y_t^2 \sum_{i,j} V_{i1} V_{i2}^{*} V_{j1} V_{j2}^{*}L_2(x_i,x_j),~~~~
A_4=Y_b^2\sum_{i,j} U_{i2} U_{j2} V_{j1} V_{i1}L_0(x_i,x_j) 
\label{Ai}
\eea
where the definition of the quantities appearing in (\ref{Ai}) 
can be found in section [2].
Notice that the renormalization scheme dependence 
in Eq.(\ref{deltaMB}) (for $\mu$ varying in the range 
$\mu \simeq (m_b/2,2m_b)$), is strongly reduced due to the NLO QCD accuracy.

As customary in this kind of analysis \cite{gluino}, 
in order to find conservative upper bounds on mass insertions,
the SM contribution to $\Delta M_{B_d}$ is set to zero. Moreover,
since we are analyzing $\Delta M_{B_d}$ which 
is a CP conserving quantity, we keep 
the squark mass matrices real. Upper bounds are then
obtained by requiring that the contribution of the real part of each 
independent combination of mass insertions in Eq.(\ref{deltaMB}) 
does not exceed the experimental central value
$\Delta M_{B_d} < 0.484 (ps)^{-1}$.

These constraints depend on the relevant MSSM low energy parameters, 
in particular, by $\tilde{m}$, $M_2$, $\mu$ and $\tan{\beta}$.\footnote{
With abuse of notation, we used here the same symbol $\mu$
for the renormalization scale of Wilson coefficients
and the Higgs mixing parameter of MSSM.}. Notice that with respect 
to the gluino mediated FCNC processes, which are
parametrized by $\tilde{m}$, $M_3$, the chargino mediated 
ones contains two free parameters more.

%%%%%%%%%%%%%%%%%%%%%%%%%%%%%%%%%%%%%%%%%%%%%%%%%%%
\begin{table}[h]
\begin{center}
\begin{tabular}{|c||c|c|c|c|}
\hline
  $M_2\;\; {\bf \backslash}\;\; m$   & 300 & 500 & 700 & 900 \\
\hline
\hline
150 & 1.3$\times 10^{-1}$ & 1.7$\times 10^{-1}$ & 2.2$\times 10^{-1}$ & 2.8$\times 10^{-1}$ \\
\hline
250 & 1.9$\times 10^{-1}$ & 2.3$\times 10^{-1}$ & 2.7$\times 10^{-1}$ & 3.2$\times 10^{-1}$ \\
\hline
350 & 2.7$\times 10^{-1}$ & 2.8$\times 10^{-1}$ & 3.3$\times 10^{-1}$ & 3.7$\times 10^{-1}$ \\
\hline
450 & 3.6$\times 10^{-1}$ & 3.6$\times 10^{-1}$ & 3.9$\times 10^{-1}$ & 4.3$\times 10^{-1}$ \\
\hline
\end{tabular}
\end{center}
\caption{Upper bounds on $\sqrt{\bigl\vert{\rm 
Re}\left[(\delta_{LL}^u)_{31}\right]^2\bigr\vert}$
from $\Delta M_{B_d}$ (assuming zero CKM and SUSY phases), 
for $\mu=200$ GeV and $\tan{\beta}=5$, and for some values 
of $\tilde{m}$ and $M_2$ (in GeV).}
\label{table2}
\end{table}
%%%%%%%%%%%%%%%%%%%%%%%%%%%%%%%%%%%%%%%%
In tables (\ref{table2}) and (\ref{table3}), 
we present our results for upper bounds on 
mass insertions coming from $\Delta M_{B_d}$, given 
for some representative values of $\tilde{m}$ and $M_2$ and 
for fixed values of $\mu=200$ GeV and $\tan{\beta}=5$.
In table (\ref{table2}) we provide constraints on
$\sqrt{\bigl\vert{\rm Re}\left[(\delta_{LL}^u)_{31}\right]^2\bigr\vert}$
for several combinations of $\tilde{m}$ and $M_2$. 
We find that these bounds are almost insensitive to $\mu$ 
and $\tan{\beta}$ in the ranges of $200-500$ GeV 
and 3--40 respectively.
This can be simply understood by noticing that the contributions to 
$\Delta B=2$ transitions mediated by LL interactions are mainly given by the 
weak gaugino component of chargino.
Therefore, the corresponding bounds are more sensitive to $M_2$ 
instead of $\mu$ and $\tan{\beta}$, since these last two parameters
contribute to the Higgsino components of chargino.
The only term in Eq.(\ref{deltaMB}) which is quite sensitive 
to $\tan{\beta}$ is $A_4$, because it is proportional to 
the bottom Yukawa coupling squared.
However, $(\delta_{LL}^u)_{31}$, in addition to $A_4$,
receives contributions also from the $A_1$ term.
This term is larger than $A_4$ and almost insensitive to $\tan{\beta}$, 
leaving the bounds on $(\delta_{LL}^u)_{31}$
almost independent from $\tan{\beta}$.

%%%%%%%%%%%%%%%%%%%%%%%%%%%%%%%%%%%%%%%%%%%%%%%%%%%%%%%%%
\begin{table}[h]
\begin{center}
\begin{tabular}{|c||c|c|c|}
\hline
m &$\sqrt{\bigl\vert{\rm Re}\left[(\delta_{LL}^u)_{31}\right]^2\bigr\vert}$  & 
$\sqrt{\bigl\vert{\rm Re}\left[(\delta_{RL}^u)_{31}\right]^2\bigr\vert}$ & 
$\sqrt{\bigl\vert{\rm Re}\left[(\delta_{LL}^u)_{31}(\delta_{LL}^u)_{32}\right]\bigr\vert}$
\\ \hline
\hline
200 & 1.4$\times 10^{-1}$ & 4.7$\times 10^{-1}$ & 2.1 $\times 10^{-1}$ \\
\hline
400 & 1.8$\times 10^{-1}$ & 9.0$\times 10^{-1}$ & 2.7$\times 10^{-1}$  \\
\hline
600 & 2.2$\times 10^{-1}$ & 1.5 & 3.4$\times 10^{-1}$ \\
\hline
800 & 2.7$\times 10^{-1}$ & 2.3 & 4.1$\times 10^{-1}$ \\
\hline
\end{tabular}
\end{center}
\caption{Upper bounds on mass insertions as in table (\ref{table2}), 
for $M_2=\mu=200$ GeV and $\tan{\beta}=5$. }
\label{table3}
\end{table}
%%%%%%%%%%%%%%%%%%%%%%%%%%%%%%%%%%%%%%%%%%%%%%%%%%%
\begin{table}[h]
\begin{center}
\begin{tabular}{|c||c|c|c|}
\hline
m & 
$\sqrt{\bigl\vert{\rm Re}\left[(\delta_{LL}^u)_{31}(\delta_{RL}^u)_{31}
\right]\bigr\vert}$ 
&
$\sqrt{\bigl\vert{\rm Re}\left[(\delta_{LL}^u)_{31}(\delta_{RL}^u)_{32}\right]\bigr\vert}$ &
$\sqrt{\bigl\vert{\rm Re}\left[(\delta_{RL}^u)_{31}(\delta_{RL}^u)_{32}\right]\bigr\vert}$ 
\\ \hline
\hline
200 & 1.8$\times 10^{-1}$ & 4.0$\times 10^{-1}$ & 7.1 $\times 10^{-1}$ \\
\hline
400 & 3.0$\times 10^{-1}$ & 6.3$\times 10^{-1}$ & 1.3 \\
\hline
600 & 4.5$\times 10^{-1}$ & 9.5$\times 10^{-1}$ & 2.3 \\
\hline
800 & 6.3$\times 10^{-1}$ & 1.3 & 3.5 \\
\hline
\end{tabular}
\end{center}
\caption{Upper bounds on mass insertions as in table (\ref{table2}), 
for $M_2=\mu=200$ GeV and $\tan{\beta}=5$. }
\label{table4}
\end{table}
%%%%%%%%%%%%%%%%%%%%%%%%%%%%%%%%%%%%%%%%%%%%%%%%%%%%%%%%%%%%%%%%%%%%%

In tables  (\ref{table3}) and (\ref{table4})  
we give our results for the real parts of the other mass insertions 
(and as well as for $\sqrt{\vert \mathrm{Re}\left[(\delta_{LL}^u)_{31}\right]^2\vert}$) 
which are less constrained, for several values of $\tilde{m}$ and evaluated at
$M_2=\mu=200$ GeV and $\tan{\beta}=5$. For larger values of $\mu$ and $M_2$, 
these bounds become clearly less stringent due to the decoupling.
Notice that they are also quite insensitive 
to $\tan{\beta}$, since no mass insertion in Eq.(\ref{deltaMB})
receives leading contributions from bottom Yukawa couplings.
It is also worth mentioning that the bounds on the mass insertion 
$(\delta^u_{LL})_{32} (\delta^u_{RL})_{31}$ are identically to the bounds
of $(\delta^u_{LL})_{31} (\delta^u_{RL})_{32}$, since they have the same 
coefficients in $C_1^{\chi}$ as can be seen from Eq.(\ref{R}). 
Therefore, here we just present the bounds of one of them.
Moreover, due to the results in Eqs.(\ref{deltaMB})--(\ref{R})
and the strategy adopted in setting constraints,
the upper bounds for the other mass insertions combinations, where 
$L \leftrightarrow R$, turn out to be exactly the same 
as the corresponding ones in 
tables (\ref{table2})--(\ref{table4}), and therefore 
we do not show them in our analysis.

In analogy to the procedure used for obtaining bounds from $\Delta M_{B_d}$,
the imaginary parts will be constrained
by switching off the SM CKM phase and 
imposing that the contribution of the SUSY phases to
$\sin{2\beta}$ does not exceed its
experimental central value $(\sin{2\beta})^{\rm exp}=0.79$. 
In particular we obtain 
\be
(\tan{2\beta})^{\rm exp} < \left(
\frac{g^4 m_{B_d}f_{B_d}^2}{(48\pi)^2\tilde{m}^2\Delta M_{B_d}}~
{\rm Im}[R] \right)
\label{Sin2b}
\ee
where $R$ is defined in Eq.(\ref{R}). 

In tables (\ref{table5})--(\ref{table7}).
we present our numerical results for the bounds on imaginary parts 
of mass insertions.
Clearly, due to the procedure used in our analysis, 
these bounds turn out to be just
proportional to the corresponding ones in tables 
(\ref{table2})--(\ref{table4}), and 
therefore the same considerations about $\mu$ and $\tan{\beta}$ dependence
hold for these bounds as well.

\begin{table}
\begin{center}
\begin{tabular}{|c||c|c|c|c|}
\hline
  $M_2\;\; { \bf \backslash}\;\; m$   & 300 & 500 & 700 & 900 \\
\hline
\hline
150 & 1.5$\times 10^{-1}$ & 2.0$\times 10^{-1}$ & 2.6$\times 10^{-1}$ & 3.1$\times 10^{-1}$ \\
\hline
250 & 2.2$\times 10^{-1}$ & 2.6$\times 10^{-1}$ & 3.1$\times 10^{-1}$ & 3.6$\times 10^{-1}$ \\
\hline
350 & 3.0$\times 10^{-1}$ & 3.3$\times 10^{-1}$ & 3.7$\times 10^{-1}$ & 4.2$\times 10^{-1}$ \\
\hline
450 & 4.0$\times 10^{-1}$ & 4.1$\times 10^{-1}$ & 4.4$\times 10^{-1}$ & 4.8$\times 10^{-1}$ \\
\hline
\end{tabular}
\end{center}
\caption{Upper bounds on $\sqrt{\bigl\vert{\rm 
Im}\left[(\delta_{LL}^u)_{31}\right]^2\bigr\vert}$
from $\sin 2\beta =0.79$ (assuming a zero CKM phase), 
for $\mu=200$ GeV and $\tan{\beta}=5$, and for some 
values of $\tilde{m}$ and $M_2$ (in GeV).}
\label{table5}
\end{table}
%%%%%%%%%%%%%%%%%%%%%%%%%%%%%%%%%%%%%%%%
\begin{table}[h]
\begin{center}
\begin{tabular}{|c||c|c|c|}
\hline
m &$\sqrt{\bigl\vert{\rm Im}\left[(\delta_{LL}^u)_{31}\right]^2\bigr\vert}$  & 
$\sqrt{\bigl\vert{\rm Im}\left[(\delta_{RL}^u)_{31}\right]^2\bigr\vert}$ & 
$\sqrt{\bigl\vert{\rm Im}\left[(\delta_{LL}^u)_{31}(\delta_{LL}^u)_{32}\right]\bigr\vert}$
\\ \hline
\hline
200 & 1.6$\times 10^{-1}$ & 5.4$\times 10^{-1}$ & 2.4 $\times 10^{-1}$ \\
\hline
400 & 2.0$\times 10^{-1}$ & 1.0 & 3.0$\times 10^{-1}$  \\
\hline
600 & 2.5$\times 10^{-1}$ & 1.7 & 3.8$\times 10^{-1}$ \\
\hline
800 & 3.1$\times 10^{-1}$ & 2.7 & 4.6$\times 10^{-1}$ \\
\hline
\end{tabular}
\end{center}
\caption{Upper bounds on mass insertions as in table (\ref{table5}),
for $M_2=\mu=200$ GeV and $\tan{\beta}=5$.}
\label{table6}
\end{table}
%%%%%%%%%%%%%%%%%%%%%%%%%%%%%%%%%%%%%%%%%%%%%%%%%%%
\begin{table}[h]
\begin{center}
\begin{tabular}{|c||c|c|c|}
\hline
m & 
$\sqrt{\bigl\vert{\rm Im}\left[(\delta_{LL}^u)_{31}(\delta_{RL}^u)_{31}
\right]\bigr\vert}$ 
&
$\sqrt{\bigl\vert{\rm Im}\left[(\delta_{LL}^u)_{31}(\delta_{RL}^u)_{32}\right]\bigr\vert}$ &
$\sqrt{\bigl\vert{\rm Im}\left[(\delta_{RL}^u)_{31}(\delta_{RL}^u)_{32}\right]\bigr\vert}$ 
\\ \hline
\hline
200 & 2.1$\times 10^{-1}$ & 4.5$\times 10^{-1}$ & 8.0 $\times 10^{-1}$ \\
\hline
400 & 3.4$\times 10^{-1}$ & 7.2$\times 10^{-1}$ & 1.5 \\
\hline
600 & 5.1$\times 10^{-1}$ & 1.1& 2.5 \\
\hline
800 & 7.2$\times 10^{-1}$ & 1.5 & 4.0 \\
\hline
\end{tabular}
\end{center}
\caption{Upper bounds on mass insertions as in table (\ref{table5}), 
for $M_2=\mu=200$ GeV and $\tan{\beta}=5$.}
\label{table7}
\end{table}
%%%%%%%%%%%%%%%%%%%%%%%%%%%%%%%%%%%%%%%%%%%%%%%%%%%%%%%%%%%%%%%%%%%%%

Next we consider the upper bounds on the relevant mass insertions 
in the down--squark sector, mediated by gluino exchange.
In Ref.\cite{gluinoB} the maximum allowed values for the real and 
imaginary parts of the mass insertions
$(\delta^d_{LL})_{13}$ and $(\delta^d_{LR})_{13}$ are given
by taking into account the NLO QCD corrections.
However, in that analysis 
the SM contributions to $\Delta M_{B_d}$ and $\sin 2 \beta$ 
are assumed not vanishing. 
In order to compare our bounds 
on up--squark mass insertions with the corresponding ones
in the down-squark sector, we should use for these last ones 
the same strategy adopted above.
Therefore, in order to find conservative upper bounds on down-squark 
mass insertions, we will impose that the pure gluino contribution does not 
exceed the experimental values on $\Delta M_{B_d}$ and $\sin 2 \beta$,
setting to zero the SM contribution.
In these results we  
include the NLO QCD corrections for Wilson coefficients given 
in (\ref{CWnlo}).

The upper bounds on the real parts of relevant combinations of
mass insertions $(\delta^d_{AB})_{13}$ (with $A,B=(L,R)$)
from the gluino contribution to 
$\Delta M_{B_d}$ are presented in table (\ref{table8}). 
In table (\ref{table9}) we show the corresponding bounds
for the imaginary parts obtained 
from the gluino contribution to CP asymmetry $a_{J/\psi K_S}$, again assuming 
zero SM contribution. The 
upper bounds on the other mass insertions in which $L\leftrightarrow R$ are 
not shown here, since, as for the chargino case,
they turn to be exactly the same as the corresponding ones in tables 
(\ref{table8}) and (\ref{table9}).

%%%%%%%%%%%%%%%%%%%%%%%%%%%%%%%%%%%%%%%%%%%%%%%%%%%%%%%%%%%%%%%%%%%%%
\begin{table}[h]
\begin{center}
\begin{tabular}{|c||c|c|c|c|}
\hline
  $M_3$   
& 
$\sqrt{\bigl\vert{\rm Re}\left[(\delta_{LL}^d)^2_{31}\right]\bigr\vert}$ 
&  
$\sqrt{\bigl\vert{\rm Re}\left[(\delta_{RL}^d)^2_{31}\right]\bigr\vert}$ 
&  
$\sqrt{\bigl\vert{\rm Re}\left[(\delta_{LL}^d)_{31}(\delta_{RR}^d)_{31}
\right]\bigr\vert}$ 
& 
$\sqrt{\bigl\vert{\rm Re}\left[(\delta_{LR}^d)_{31}(\delta_{RL}^d)_{31}
\right]\bigr\vert}$ 
\\
\hline
\hline
200 & 4.6$\times 10^{-2}$ & 2.2$\times 10^{-2}$ & 8.4$\times 10^{-3}$ 
& 1.1$\times 10^{-2}$ \\
\hline
400 & 1.0$\times 10^{-1}$ & 2.4$\times 10^{-2}$ & 9.6$\times 10^{-3}$ 
& 1.9$\times 10^{-2}$ \\
\hline
600 & 4.8$\times 10^{-1}$ & 2.9$\times 10^{-2}$ & 1.2$\times 10^{-2}$ 
& 3.0$\times 10^{-2}$ \\
\hline
800 & 2.4$\times 10^{-1}$ & 3.4$\times 10^{-2}$ & 1.4$\times 10^{-2}$ 
& 4.4$\times 10^{-2}$ \\
\hline
\end{tabular}
\end{center}
\caption{Upper bounds on real parts of combinations of 
mass insertions $(\delta_{AB}^d)_{31}$, with 
$(A,B)=L,R$, from gluino contributions
to $\Delta M_{B_d}$ (assuming zero SM contribution), 
evaluated at $\tilde{m}=400$ GeV and for some values of 
gluino mass $M_3$ (in GeV).} 
\label{table8}
\end{table}
%%%%%%%%%%%%%%%%%%%%%%%%%%%%%%%%%%%%%%%%%%%%%%%%%%%%

%%%%%%%%%%%%%%%%%%%%%%%%%%%%%%%%%%%%%%%%%%%%%%%%%%%%%%%%%%%%%%%%%%%%%
\begin{table}[h]
\begin{center}
\begin{tabular}{|c||c|c|c|c|}
\hline
  $M_3$   
& 
$\sqrt{\bigl\vert{\rm Im}\left[(\delta_{LL}^d)^2_{31}\right]\bigr\vert}$ 
&  
$\sqrt{\bigl\vert{\rm Im}\left[(\delta_{RL}^d)^2_{31}\right]\bigr\vert}$ 
&  
$\sqrt{\bigl\vert{\rm Im}\left[(\delta_{LL}^d)_{31}(\delta_{RR}^d)_{31}
\right]\bigr\vert}$ 
& 
$\sqrt{\bigl\vert{\rm Im}\left[(\delta_{LR}^d)_{31}(\delta_{RL}^d)_{31}
\right]\bigr\vert}$ 
\\
\hline
\hline
200 & 5.2$\times 10^{-2}$ & 2.5$\times 10^{-2}$ & 9.6$\times 10^{-3}$ 
& 1.2$\times 10^{-2}$ \\
\hline
400 & 1.2$\times 10^{-1}$ & 2.7$\times 10^{-2}$ & 1.1$\times 10^{-2}$ 
& 2.2$\times 10^{-2}$ \\
\hline
600 & 5.5$\times 10^{-1}$ & 3.3$\times 10^{-2}$ & 1.3$\times 10^{-2}$ 
& 3.4$\times 10^{-2}$ \\
\hline
800 & 2.8$\times 10^{-1}$ & 3.9$\times 10^{-2}$ & 1.6$\times 10^{-2}$ 
& 5.0$\times 10^{-2}$ \\
\hline
\end{tabular}
\end{center}
\caption{Upper bounds on imaginary parts of combinations 
mass insertions $(\delta_{AB}^d)_{31}$, with 
$(A,B)=L,R$, from gluino contributions
to $\sin{2\beta}$ (assuming zero SM contribution), 
evaluated at $\tilde{m}=400$ GeV
and for some values of gluino mass $M_3$ (in GeV).}
\label{table9}
\end{table}
%%%%%%%%%%%%%%%%%%%%%%%%%%%%%%%%%%%%%%%%%%%%%%%%%%%%
\section{{\large \bf Light stop scenario}}
In this section we will provide analytical and numerical 
results for the bounds on mass insertions,
in the particular case in which one of the eigenvalues of the up--squark 
mass matrix is much lighter than the other (almost degenerates) 
ones. This scenario appears in the specific model
that we will analyze in section (5.3), 
where the mass of the stop--right ($m^2_{\tilde{t}_R}$)
is lighter than the other diagonal terms 
in the up--squark mass matrix.
Then the analytical results for the Wilson coefficients 
provided in section (3) will be generalized by including this 
effect.\footnote{We do not consider here the 
contributions of a light right-stop
to the $\tilde{Q}_i$ operators, since in this case the effect
of two mass insertions $(\Delta_{RR}^d)_{31}$ can invalidate the 
MIA method, being no heavy squarks running in the loop.}
In our case, this modification will affect only the expression for the 
Wilson coefficient $C^{\chi}_1(M_S)$ in Eq.(\ref{C1}), since the stop--right
does not contribute to $C^{\chi}_3(M_S)$ at ${\cal O}(\lambda)$ order,
as it can be seen from Eq.(\ref{C3}). 

By taking different the mass of the stop--right from the 
average squark mass, we obtain the following result\footnote{
We have used the same method introduced in Ref.\cite{BRS}, but
our results are presented in a different way.}
\bea
&&C_1^{\chi}(M_S) = \frac{g^4}{768 \pi^2 \tilde{m}^2} 
\sum_{i,j}
\Big\{\vert V_{i1}\vert^2 \vert V_{j1} \vert^2\left( 
(\delta^u_{LL})_{31}^2 + 2\lambda (\delta^u_{LL})_{31}
(\delta^u_{LL})_{32}\right) L_2(x_i,x_j)
\nonumber \\ 
&-& 2 Y_t  
\vert V_{i1}\vert^2 V_{j1} V_{j2}^*\Big(
(\delta^u_{LL})_{31} (\delta^u_{RL})_{31}+\lambda
(\delta^u_{LL})_{32} (\delta^u_{RL})_{31}+\lambda
(\delta^u_{LL})_{31} (\delta^u_{RL})_{32}\Big) R_2(x_i,x_j,z)
\nonumber \\
&+& Y_t^2 V_{i1} V_{i2}^{*} V_{j1} V_{j2}^{*}\Big(
(\delta^u_{RL})_{31}^2 +2\lambda
(\delta^u_{RL})_{31} (\delta^u_{RL})_{32}~ \Big) \tilde{R}_2(x_i,x_j,z)
\, \Big\},
\label{C1stop}
\eea
where 
$x_i= m^2_{\tilde{\chi}_i^+}/\tilde{m}^2$, $z=m^2_{\tilde{t}_R}/\tilde{m}^2$
and the functions $R_2(x,y,z)$ and $\tilde{R}_2(x,y,z)$ are given by 
\bea
R_2(x,y,z)&=&\frac{1}{x-y}\left(H_2(x,z)-H_2(y,z)\right),~~~
\tilde{R}_2(x,y,z)=\frac{1}{x-y}\left(\tilde{H}_2(x,z)-
\tilde{H}_2(y,z)\right)
\nonumber \\
H_2(x,z)&=&\frac{3}{D_2(x,z)} \Big\{ (-1 + x) ( x - z)( -1 + z) 
( -1 - x - z + 3 x z )\nonumber\\
&+& 6 x^2 ( -1 + z)^3 \log(x) - 6 ( -1 + x)^3 z^2 \log (z) \Big\}
\nonumber \\
\tilde{H}_2(x,z)&=&\frac{-6}{\tilde{D}_2(x,z)} \Big\{ (-1+x) ( x - z) (-1 + z)
( x + ( -2 + x ) z) \nonumber\\
&+& 6 x^2 ( -1 + z )^3 \log (x) - 6 (-1+x)^2 z (-2 x + z + z^2) \log (z) \Big\}
\eea
where
$D_2(x,z)={{{\left( -1+x \right) }^3}\,
     \left( x - z \right) \,{{\left(-1 + z \right) }^3}}$
and
$\tilde{D}_2(x,z)={{\left( -1 + x\right) }^2}\,
     {{\left( x - z \right) }^2}\,{{\left( -1 + z \right) }^3}$.
Notice that in the limit $z\to 1$, both the functions 
$R_2(x,y,z)$ and $\tilde{R}_2(x,y,z)$ tend to $L_2(x,y)$, recovering the
result in Eq.(\ref{C1}).
Analogously, the expressions for
$A_2$ and $A_3$ entering in Eq.(\ref{R}) must be substituted by
\be
A_2=Y_t  
\sum_{i,j}\vert V_{i1}\vert^2 V_{j1} V_{j2}^*R_2(x_i,x_j,z),~~~
A_3=Y_t^2 \sum_{i,j} V_{i1} V_{i2}^{*} V_{j1} V_{j2}^{*}\tilde{R}_2(x_i,x_j,z)
\ee
while $A_1$ and $A_4$ remain the same. In tables (\ref{table10}) and 
(\ref{table11}) we show our results, analogous
to the ones in tables (\ref{table3})-(\ref{table6}), for the
bounds on real and imaginary parts on mass insertions respectively,
by taking into account a light stop--right mass. We considered two
representative cases of $\tilde{m}_{t_R}=100, ~200$ GeV.
Clearly, the light stop--right effect does not affect bounds on mass insertions
containing LL interactions. 
From these results we could see that the effect of taking 
$\tilde{m}_{t_R} < \tilde{m}$ is sizable. In particular, on the bounds 
of the mass insertions $(\delta_{RL}^u)_{31}(\delta_{RL}^u)_{3i}$,
$(i=1,2)$ which are the most sensitive to a light stop--right.

From the results in tables (10) and (11), it is remarkable 
to notice that,
in the limit of very heavy squark masses but with fixed right stop and 
chargino masses, the bounds on $(\delta^u_{RL})_{31} 
(\delta^u_{RL})_{3i}$ tend to constant values. This is indeed an interesting 
property which shows a particular non--decoupling 
effect of supersymmetry when two 
light right--stop run inside the diagrams in Fig. (1).  This feature is 
related to the infrared singularity of the loop function 
$\tilde{R}_2(x,x,z)$ in the limit $z \to 0$. 
In particular, we find that $\lim_{z\to 0} \tilde{R}_2(x,x,z)=f(x)/x$, where
$x=m_{\chi}^2/\tilde{m}^2$, and $f(x)$ is a non-singular and non-null function
in $x=0$. Then, in the limit $\tilde{m} >> m_{\chi}$ the
 rescaling factor $1/\tilde{m}^2$ in $C_1^{\chi}$ will be canceled by the 
$1/x$ dependence in the loop function and replaced by $1/m_{\chi}^2$
times a constant factor.

This is a quite interesting result, since it shows that in the
case of light right stop and charginos masses, in comparison to the other 
squark masses, the SUSY contribution (mediated by charginos) to the 
$\Delta B=2$ processes might not decouple and could be sizable, 
provided that the mass insertions $(\delta^u_{RL})_{3i}$ are large enough.
This effect could be achieved, for instance, in supersymmetric models with
non--universal soft breaking terms. 

\begin{table}[h]
\begin{center}
\begin{tabular}{|c|c||c|c|c|}
\hline
$\tilde{m}$
& 
$\tilde{m}_{t_R}$
& 
$\sqrt{\bigl\vert{\rm Re}\left[(\delta_{RL}^u)^2_{31}\right]\bigr\vert}$ 
&  
$\sqrt{\bigl\vert{\rm Re}
\left[(\delta_{LL}^u)_{31}(\delta_{RL}^u)_{3i}\right]\bigr\vert}$ 
& 
$\sqrt{\bigl\vert{\rm Re}\left[(\delta_{RL}^u)_{31}(\delta_{RL}^u)_{32}
\right]\bigr\vert}$ 
\\
\hline
\hline
400 & 100 & 1.9$\times 10^{-1}$ & 1.6(3.3)$\times 10^{-1}$ & 2.8$\times 10^{-1}$ \\
\hline
600 & 100 & 1.8$\times 10^{-1}$ & 1.9(4.0)$\times 10^{-1}$ & 2.6$\times 10^{-1}$ \\
\hline
800 & 100 & 1.8$\times 10^{-1}$ & 2.3(4.9)$\times 10^{-1}$ & 2.6$\times 10^{-1}$ \\
\hline
\hline
400 & 200 & 3.5$\times 10^{-1}$ & 2.0(4.2)$\times 10^{-1}$ & 5.2$\times 10^{-1}$ \\
\hline
600 & 200 & 3.3$\times 10^{-1}$ & 2.3(5.0)$\times 10^{-1}$ & 4.9$\times 10^{-1}$ \\
\hline
800 & 200 & 3.2$\times 10^{-1}$ & 2.8(5.9)$\times 10^{-1}$ & 4.8$\times 10^{-1}$\\
\hline
\end{tabular}
\end{center}
\caption{Upper bounds on real parts of mass insertions 
as in tables (\ref{table3})--(\ref{table4}), for some values 
of $\tilde{m}$ and $\tilde{m}_{t_R}$ (in GeV). 
In the fourth column the first number and the one in parenthesis correspond 
to $i=1$ and $i=2$ respectively.
Upper bounds on mass insertions involving only LL interactions
are the same as in tables (\ref{table3})--(\ref{table4}).}
\label{table10}
\end{table}
%%%%%%%%%%%%%%%%%%%%%%%%%%%%%%%%%%%%%%%%%%%%%%%%%%%%

%%%%%%%%%%%%%%%%%%%%%%%%%%%%%%%%%%%%%%%%%%%%%%%%%%%%%%%%%%%%%%%%%%%%%
\begin{table}[h]
\begin{center}
\begin{tabular}{|c|c||c|c|c|}
\hline
$\tilde{m}$
& 
$\tilde{m}_{t_R}$
& 
$\sqrt{\bigl\vert{\rm Im}\left[(\delta_{RL}^u)^2_{31}\right]\bigr\vert}$ 
&  
$\sqrt{\bigl\vert{\rm Im}
\left[(\delta_{LL}^u)_{31}(\delta_{RL}^u)_{3i}\right]\bigr\vert}$ 
& 
$\sqrt{\bigl\vert{\rm Im}\left[(\delta_{RL}^u)_{31}(\delta_{RL}^u)_{32}
\right]\bigr\vert}$ 
\\
\hline
\hline
400 & 100 & 2.1$\times 10^{-1}$ & 1.8(3.7)$\times 10^{-1}$ & 3.1$\times 10^{-1}$ \\
\hline
600 & 100 & 2.0$\times 10^{-1}$ & 2.2(4.6)$\times 10^{-1}$ & 3.0$\times 10^{-1}$ \\
\hline
800 & 100 & 2.0$\times 10^{-1}$ & 2.6(5.5)$\times 10^{-1}$ & 3.0$\times 10^{-1}$ \\
\hline
\hline
400 & 200 & 4.0$\times 10^{-1}$ & 2.2(4.8)$\times 10^{-1}$ & 6.0$\times 10^{-1}$ \\
\hline
600 & 200 & 3.7$\times 10^{-1}$ & 2.7(5.6)$\times 10^{-1}$ & 5.6$\times 10^{-1}$ \\
\hline
800 & 200 & 3.6$\times 10^{-1}$ & 3.1(6.7)$\times 10^{-1}$ & 5.4$\times 10^{-1}$\\
\hline
\end{tabular}
\end{center}
\caption{
Upper bounds on imaginary parts of mass insertions 
as in tables (\ref{table5})--(\ref{table6}), for some values
of $\tilde{m}$ and $\tilde{m}_{t_R}$ $\tilde{m}$ (in GeV).
In the fourth column the first number and the one in parenthesis correspond 
to $i=1$ and $i=2$ respectively.
Upper bounds on mass insertions involving only LL interactions
are the same as in tables (\ref{table5})--(\ref{table6}).}
\label{table11}
\end{table}
%%%%%%%%%%%%%%%%%%%%%%%%%%%%%%%%%%%%%%%%%%%%%%%%%%%%
\section{{\large \bf Specific supersymmetric models}}

In this section we focus on three specific supersymmetric models and study 
the impact of the constraints derived 
in previous sections on their predictions.
We discuss first SUSY models with minimal flavor violation, then we study 
the ones with Hermitian flavor structure, and finally 
we consider a SUSY model with 
small CP violating phases with universal strength of Yukawa couplings.
%%%%%%%%%%%%%%%%%%%%%

\subsection{{\large \bf SUSY models with minimal flavor violation}}

In supersymmetric models with minimal flavor violation (MFV) 
the CKM matrix is the only source of flavor violation.
In the framework of MSSM (with $R$ parity conserved)
we consider a minimal model, like in the supergravity scenario,
where the soft SUSY 
breaking terms is assumed to be universal at 
grand unification scale, {\it, i.e.},
the soft scalar masses, gaugino masses and 
trilinear and bilinear couplings are given by
\be
m_i^2 = m_0^2, ~~~~ M_a = m_{1/2} e^{-i\alpha_M}, ~~~~ A_{\alpha} = A_0 
e^{-i \alpha_A}, ~~~~ B= B_0 e^{-i\alpha_B}.
\ee
As mentioned in the introduction, only two of the above phases are independent,
and can be chosen as 
\be 
\phi_A = \mathrm{arg} ( A^* M), ~~~~~~~~ \phi_B = \mathrm{arg} (B^* M).
\ee
The main constraints on $\phi_A$ and $\phi_B$ are due 
to the EDM of the electron,
neutron and mercury atom. The present experimental bound on EDMs implies that 
$\phi_{A,B}$ should be $\lsim 10^{-2}$ unless the SUSY 
masses are unnaturally heavy\cite{phase:edm}.

In these scenarios, where SUSY phases $\phi_{A,B}$ are 
constrained to be very small by EDMs bounds, the supersymmetric contributions
to CP violating phenomena in $K$ and $B$ mesons
do not generate any sizable deviation from the SM prediction. 
We have to mention 
that the universal structure for the soft breaking terms, 
specially the universality 
of the trilinear couplings, is a very strong assumption. Indeed, in the 
light of  
recent works on SUSY breaking in string theories, the soft breaking sector
at GUT scale is generally found to be 
non--universal \cite{ibanez}.
Notice that, even if we start with universal 
soft breaking terms at GUT scale, some off diagonal terms in the squark mass 
matrices are induced at electroweak (EW) scale by Yukawas interactions 
through the renormalization group equation (RGE) evolution.
Therefore these off--diagonal entries are suppressed
by the smallness of the CKM angles and/or the smallness of the 
Yukawa couplings.

It is important to stress
that even though one ignores the bounds from the EDMs and 
allows larger values (of order ${\cal O}(1)$ ) for the 
SUSY phases $\phi_{A,B}$, 
this class of models with MFV can not generate any large
contribution to $\varepsilon_K$  
and $\varepsilon'/\varepsilon$. Therefore, the Yukawa 
couplings remain the main source of CP violation \cite{barr}. 

Here we also found that, within MFV scenarios, the 
SUSY contributions to $\Delta M_{B_d}$ and $a_{J/\psi K_S}$ are negligible. 
In fact, due to the universality assumption of soft 
SUSY breaking terms, it turns out that the gluino and 
chargino contributions are quite suppressed. 
For instance, for $m_0 \sim m_{1/2} \sim A_0 
\sim 200$ GeV and $\phi_{A,B} \sim \pi/2$ (which corresponds to
 $\tilde{m}^2$ and $m_g$ at SUSY scale of order
$500$ GeV) we find the following values of the relevant mass insertions: 
$\mathrm{Im}(\delta^d_{13})_{LL} \sim \mathrm{Re}(\delta^d_{13})_{LL} 
\sim 10^{-4}$ and $\mathrm{Im}(\delta^d_{13})_{LR} \sim \mathrm{Re}
(\delta^d_{13})_{LR} \sim 10^{-6}$, which are clearly much smaller than 
the corresponding bounds mentioned in 
the previous section.\footnote{In our analysis we have taken into account the
effect of the CP violating phases in the RGE evolution.}
 
Therefore, we conclude that SUSY models with MFV do not give any 
genuine contribution to the CP violating 
and flavor changing processes in $K$ and $B$ systems and 
this scenario can not be distinguished from the SM model one.

%%%%%%%%%%%%%%%%%%%%%%%%%%%%%%%%%%%%%%%%%%%%%%%%%%%%%%%%%%%%%

\subsection{{\large \bf SUSY models with Hermitian flavor structure}}

As discussed in the introduction, a possible solution for suppressing 
the EDMs in
SUSY model is to have Hermitian flavor structures \cite{hermitian}. 
In this class 
of models, the flavor blind quantities, such as the $\mu$--terms and 
gaugino masses,
are real while the Yukawa couplings and $A$-terms are Hermitian, i.e 
$Y_{u,d}^{\dagger}= Y_{u,d}$ and $A_{u,d}^{\dagger}= A_{u,d}$.  It has been 
shown that these models are free from the EDM constraints and the 
off--diagonal 
phases lead to significant contribution to the observed CP violation
in the kaon system, in particular to 
$\varepsilon'/\varepsilon$ \cite{hermitian}.

Let us consider, for instance, the case of Hermitian and hierarchical 
quark mass matrices with three zeros \cite{Froggatt}
\begin{equation}
M_i = \left( \begin{array}{ccc}
0 & a_i e^{i\alpha_i}& 0 \\
a_i e^{-i\alpha_i} & A_i & b_i e^{i\beta_i} \\
0 & b_i e^{-i\beta_i} & B_i
\end{array} \right) \;;~~~ i=u,d
\end{equation}
with $A_i=(m_c, m_s)$, $B_i = (m_t-m_u, m_b-m_d)$, 
$a_i= (\sqrt{m_u m_c},\sqrt{m_d m_s})$,
and $b_i = (\sqrt{m_u m_t},\sqrt{m_d m_b})$. The phases $\alpha_i$ and 
$\beta_i$ satisfy: 
$\alpha_d - \alpha_u = \pi/2$ and $\beta_d - \beta_u = \pi/2$. 
These matrices reproduce the correct values for 
quark masses and CKM matrix. 
We also assume the following Hermitian $A$--terms:
\begin{equation}
A_d=A_u = \left( \begin{array}{ccc}
A_{11} & A_{12}~e^{i\varphi_{12}} & A_{13}~e^{i\varphi_{13}}  \\
A_{12}~e^{-i\varphi_{12}} & A_{22} &A_{23}~e^{i\varphi_{23}}  \\
A_{13}~e^{-i\varphi_{13}}&A_{23}~e^{-i\varphi_{23}}  & A_{33}
\end{array} \right) \;.
\end{equation}
Notice that, the scenario with non--degenerate $A$-terms 
is an interesting possibility for enhancing the 
SUSY contributions to $\varepsilon_K$ and $\varepsilon'/\varepsilon$ 
\cite{susycp} and it is also well motivated by many string inspired models. 
In this case, the mass insertions are given by
\bea
(\delta^{q}_{ij})_{LL} &=&   \frac{1}{m_{\tilde{q}}^2} \left(V^{q} M_Q^2 
V^{q^{\dagger}}\right)_{ij}\\
(\delta^{q}_{ij})_{LR} &=& \frac{1}{m_{\tilde{q}}^2}
\left[ \left(V^{q} Y^{A^*}_q V^{q^{\dagger}}\right)_{ij} v_{1(2)} -
\mu Y^{q}_i \delta_{ij} v_2(1) \right],
\eea
where $q\equiv u,d$ and $(Y^A_q)_{ij}=Y^q_{ij} A^q_{ij}$. 
Since the Yukawa are Hermitian matrices, they are diagonalized 
by only one unitary transformation. 

In this class of models, we find that in most of the parameter space 
the chargino gives the dominant contribution 
to $B_d - \bar{B}_d$ mixing and CP 
asymmetry $a_{J/\psi K_S}$, while the gluino one is sub-leading. 
As we emphasized above, 
in order to have a significant gluino contribution for $\tilde{m}\sim m_g\sim
500$ GeV ({\it i.e.}, $m_0\sim M_{1/2}\sim 200$ at GUT scale), the real and 
imaginary parts of mass insertion 
$(\delta^d_{13})_{LL}$ or $(\delta^d_{13})_{LR}$ 
should be of order $10^{-1}$ and $10^{-2}$ respectively. 
However, with the above hierarchical Yukawas we find 
that these mass insertions are two orders of magnitude below the required 
values so that the gluino contributions are very small.

Concerning the chargino amplitude to the CP asymmetry $a_{J/\psi K_S}$,
we find that the mass insertions 
$(\delta^u_{31})_{RL}$ and $(\delta^u_{31})_{LL}$
give the leading contribution to $a_{J/\psi K_S}$.
 However, for 
the representative case of $m_0 = m_{1/2} =200$ and $\phi_{ij}
\simeq \pi/2$ the values of these mass insertions are given by
\begin{eqnarray}
\sqrt{\vert \mathrm{Im}[(\delta^u_{LL})_{31}]^2\vert}&=&6\times 10^{-4},\\
\sqrt{\vert \mathrm{Im}[(\delta^u_{LL})_{31}]^2\vert}&=&4\times 10^{-3},\\
\sqrt{\vert \mathrm{Im}[(\delta^u_{LL})_{31} (\delta^u_{RL})_{32}]\vert}&=&
1\times 10^{-4}.
\end{eqnarray}
These results show that, also for this class of models,
SUSY contributions cannot give sizable 
effects to $a_{J/\psi K_S}$.  
As expected, with hierarchical Yukawa couplings 
(where the mixing between different generations is very small), 
the SUSY contributions to the $B-\bar{B}$ mixing and the CP asymmetry of 
$B_d \to J/\psi K_S$ are sub-dominant and the SM should give the dominant 
contribution.

%%%%%%%%%%%%%%%%%%%%%%%%%%%%%%%%%%%%%%%%%%%%%%%%%%%%%%%%%%%
\subsection{{\large \bf SUSY model with universal strength of Yukawa couplings}}

Supersymmetric models with small CP violating 
phases is a possible solution for suppressing the EDMs.
In Ref.\cite{branco} it was shown that, among this class of models, 
the ones with universal strength of Yukawa couplings naturally 
provide very small CP violating phases.
However due to the large mixing between different generations, 
it was found that the $LL$ mass insertions 
can give sizable effects to $\varepsilon_K$ and 
$\varepsilon'/\varepsilon$ by means of gluino and chargino 
exchanges respectively.
Furthermore, it was also emphasized that 
in these models the SM contribution to 
the CP asymmetry $a_{J/\psi K_S}$ might be negligible, leaving the dominant
SUSY effect (due to the chargino exchange) to 
account for the experimental results.

Here we will discuss the different contributions to
$B_d-\bar{B}_d$ and $a_{J/\psi K_S}$ in terms of mass insertions and 
compare the predictions of this model with the corresponding ones 
of Hermitian flavor structure discussed in the previous subsection. 
In the framework
of the universal strength Yukawa couplings, the quark Yukawa couplings can be
written as
\begin{equation}
U_{ij} = \frac{\lambda_u}{3} \exp \left[i \Phi^u_{ij}\right] \quad
\mathrm{and} \quad D_{ij} = \frac{\lambda_d}{3} \exp \left[i
\Phi^d_{ij}\right] \;,
\label{usy}
\end{equation}
where $\lambda_{u,d}$ are overall real constants, and 
$\Phi^{u,d}$ are pure phase matrices which are constrained to be very small by
the hierarchy of the quark masses \cite{branco}. 
The values of these parameters, that
lead to the correct quark spectrum and mixing, can be found in 
Ref.\cite{branco}. 
As explained in that paper, an important feature 
of this model is the presence of a large mixing 
between the first and third generation.
As we will show in the following, this property will 
account for large SUSY contributions to $a_{J/\psi K_S}$. 

%%%%%%%%%%%%%%%%%%%%%%%%%%%%%%%%%%%%%%%% 
In the framework of universal strength Yukawa couplings Eq.(\ref{usy}), due
to the large generation mixing, the EDMs impose severe constraints on the 
parameter space and force the trilinear couplings to take particular patterns
as the factorizable matrix form~\cite{branco}, {\it i.e.},
\be
A = m_0 \left( \begin{array}{ccc}
a & a & a \\
b & b & b \\
c & c & c
\end{array} \right) .
\ee
In order to satisfy the bound of the mercury EDM, the phases of the entries
$a,b$ and $c$ should be of order $10^{-2}-10^{-1}$ ~\cite{branco}.
As an illustrative example, we consider $m_0 = m_{1/2}= 200$ GeV, $\phi_a=
\phi_c=0, \phi_b=0.1$  and $a=-1, b=-2, c=-3$. In this case one 
finds that at low energy
the average squark mass is of order 500 GeV, however
one of the stop masses ($\tilde{t}_R$) is much lighter, 
$m_{\tilde{t}_R}\simeq 200$ GeV. 
The gaugino mass $M_2$ is of order 170 GeV and, 
from the EW breaking condition, $\vert \mu \vert$ turns out 
to be of order of 400 GeV. In this case, the relevant mass 
insertions for the gluino contribution are given by
\be
(\delta^d_{LL})_{31} \simeq - 0.001+ 0.02~ i,~~~~~~~~~~~
(\delta^d_{RL})_{31}\simeq  0.00002+ 0.0009~ i.
\ee
Regarding the other mass insertions (LR and RR), they
are much smaller ($\lsim {\cal O}(10^{-6})$), and so 
we do not show them. It is clear that, with these values 
for the down--squark mass
insertions, the gluino contribution to $a_{J/\psi K_S}$ is negligible 
(of the order of $3\%$). 

On the contrary, the relevant up--squark mass insertions 
for the chargino contribution are given by
\bea
(\delta^u_{LL})_{31} &\simeq& 0.001 + 0.05~ i,~~~~~~~~~~~
(\delta^u_{RL})_{31} \simeq -0.0004+ 0.13~ i,\\
(\delta^u_{LL})_{32} &\simeq & -0.008 -0.11~ i,~~~~~~~~~~~
(\delta^u_{RL})_{32} \simeq 0.01 - 0.28~i.
\eea
Comparing these results with the ones in tables (\ref{table10}) and
(\ref{table11}), we see that for this model, the chargino contribution 
to the imaginary parts $(\delta^u_{RL})_{31}$ and 
$(\delta^u_{RL})_{32}$ is of the same order of the corresponding upper bounds.
Notice that these imaginary parts are of the same order, 
so that they might coherently contribute 
to give a sizable effect on $a_{J/\psi K_S}$. 
In particular, by using the exact 1-loop calculation, we find that 
the chargino contribution leads to $\sin (2 \theta_d) \sim 0.75$.
Moreover, as a check on our computations, 
we have compared our results from MIA approximation with the 
corresponding ones obtained by using the full
calculation \cite{branco}. 
In this case we find that, by taking into account the 
effect of a light stop, the MIA predictions are quite compatible with the 
results of the full computation.

\section{Conclusions}
In this paper we have studied the chargino contributions to $B_d-\bar{B}_d$ 
mixing and CP asymmetry $a_{J/\psi K_S}$ in the mass insertion 
approximation. In our analysis we have taken into account the NLO QCD 
corrections to the effective Hamiltonian for $\Delta B=2$ transitions
$H_{eff}^{\Delta B=2}$. We provided analytical results for the 
chargino contribution to $H_{eff}^{\Delta B=2}$ in the 
framework of mass insertion method, and 
given the expressions for the 
$B_d-\bar{B}_d$ and CP asymmetry $a_{J/\psi K_S}$ at NLO in QCD, 
as a function of mass insertions in the up--squark sector.
We have also provided model independent upper bounds on mass insertions 
by requiring that the pure chargino contribution
does not exceed the experimental values of $B-\bar{B}$ mixing and 
CP asymmetry $a_{J/\psi K_S}$.
Since in many SUSY models the chargino contribution gives the dominant effect
to $B-\bar{B}$ mixing and CP asymmetry $a_{J/\psi K_S}$, our results
are particularly useful for a ready check of the viability of 
these models.
Moreover we generalized our results by including the case of a 
light right-stop scenario. In this case we found the interesting property 
that the bounds on mass insertions combinations
$(\delta_{RL}^u)_{31}(\delta_{RL}^u)_{3i}$ are not sensitive
to the common squark mass when this is very large in comparison to 
the chargino and stop--right ones.

Finally, we applied these results to 
a general class of SUSY models which are particularly suitable to 
solve the SUSY CP problem, namely the SUSY models with minimal 
flavour violations, hermitian flavour structure, and small CP 
violating phases with universal strength Yukawa couplings. 
We have shown that in SUSY models
with minimal flavor violation and with Hermitian (and hierarchical) Yukawa
couplings and $A$--terms, the SUSY contributions to the $B-\bar{B}$ mixing 
and the CP asymmetry $a_{\psi K_S}$ are very small and the SM contribution 
in these classes of models should give the dominant effect. 
On the contrary, 
in the case of SUSY scenarios with
large mixing between different generations in the soft terms,
the SUSY contributions become significant and 
can even be the dominant source for 
saturating the experimental value of 
$a_{\psi K_S}$. Among this class of models, 
we have investigated a SUSY model with universal strength of 
Yukawa couplings. In this case, we have found that
the chargino exchange provides the leading contribution to 
$a_{\psi K_S}$ through the mass insertions 
$(\delta_{LL}^u)_{31}(\delta_{RL}^u)_{3i}$, $i=1,2$ and 
$(\delta_{RL}^u)_{31}(\delta_{RL}^u)_{32}$.

%%%%%%%%%%%%%%%%%%%%%%%%%%%%%%%%%%%%%%%%%%%%%%%%%%%%%%%%%%%%
%
\section*{\bf \normalsize Acknowledgements}
We acknowledge the kind hospitality of the CERN Theory Division
where part of this work has been done. The work of S.K. was supported by PPARC.
E.G. would like to thank Katri Huitu for useful discussions.

%%%%%%%%%%%%%%%%%%%%%%%%%%%%%%%%%%%%%%%%%%%%%%%%%%%
%

%%%%%%%%%%%%%%%%%%%%%%%%%%%%%%%%%%%%%%%%%%%%%%%%%%%%%%%%%%%%%%%%%%%%%
\end{document}